# Effects of hydrogen/deuterium absorption on the magnetic properties of Co/Pd multilayers

K. Munbodh*, F. A. Perez, C. Keenan, and D. Lederman

*Department of Physics, West Virginia University, Morgantown, WV 26506*

M. Zhernenkov and M.R. Fitzsimmons

*Los Alamos National Laboratory, Los Alamos, NM 87545*

## ABSTRACT

The effects of hydrogen ($H_2$) and deuterium ($D_2$) absorption were studied in two Co/Pd multilayers with perpendicular magnetic anisotropy (PMA) using polarized neutron reflectivity (PNR). PNR was measured in an external magnetic field $H$ applied in the plane of the sample with the magnetization $M$ confined in the plane for $\mu_o H = 6.0$ T and partially out of plane at 0.65 T. Nominal thicknesses of the Co and Pd layers were 2.5 Å and 21 Å, respectively. Because of these small values, the actual layer chemical composition, thickness, and interface roughness parameters were determined from the nuclear scattering length density profile ($\rho_n$) and its derivative obtained from both x-ray reflectivity and PNR, and uncertainties were determined using Monte Carlo analysis. The PNR $\rho_n$ showed that although $D_2$ absorption occurred throughout the samples, absorption in the multilayer stack was modest (0.02 D per Pd atom) and thus did not expand. Direct magnetometry showed that $H_2$ absorption decreased the total $M$ at saturation and increased the component of $M$ in the plane of the sample when not at saturation. The PNR magnetic scattering length density ($\rho_m$) revealed that the Pd layers in the multilayer stack were magnetized and that their magnetization was preferentially modified upon $D_2$ absorption. In one sample, a modulation of $M$ with twice the multilayer period was observed at $\mu_o H =$ 0.65 T, which increased upon $D_2$ absorption. These results indicate that $H_2$ or $D_2$ absorption decreases both the PMA and total magnetization of the samples. The lack of measurable expansion during absorption indicates that these changes are primarily governed by modification of the electronic structure of the material.

*Corresponding author. e-mail address: kmunbodh@mix.wvu.edu

## I. INTRODUCTION

$H_2$ absorption in Pd-based thin film structures has recently attracted significant interest due to their ability to store and release large quantities of $H_2$ at room temperature.[1-3] When a $H_2$ molecule is adsorbed on the surface of bulk Pd, it dissociates into two H atoms which diffuse into the Pd lattice.[4] At room temperature, there are two phases of PdH, designated as $\alpha$ and $\beta$ phases. When the concentration of H is greater than 60% ($\beta$ phase), the lattice parameter increases up to 3.6%.[4]

$H_2$ interactions with metallic thin films and multilayers can significantly modify their electronic, magnetic, optical, and structural properties.[5-8] In particular, magnetic coupling between ferromagnetic thin layers mediated by non-magnetic layers is influenced by $H_2$ absorption. For example, magnetization and neutron reflectivity measurements have shown that in Fe/Nb multilayers magnetic coupling between Fe layers switches from antiferromagnetic to ferromagnetic upon $H_2$ absorption. This has been attributed to a change in the effective Fermi wavevector in the Nb layers, which changes the sign of the electronic Ruderman-Kittel-Kasuya-Yosida (RKKY) interaction responsible for coupling between Fe layers.[9] In Fe/V multilayers, changes in their magnetic properties result from a re-distribution of the Fe and V *d*-electrons at the interfaces.[10-12]

In Pd/Co/Pd trilayers, perpendicular magnetic anisotropy (PMA) initially increases and then decreases with time upon $H_2$ absorption as a result of a modification of the magnetic properties of ultra-thin Co films induced by H in surrounding Pd layers.[13] However, it is unclear what the effects of $H_2$ absorption are on other possible mechanisms that may affect magnetic properties, such as magnetoelastic coupling, which are known to be important in Co/Pd multilayers.[14, 15]

To understand the mechanisms responsible for the modifications of Pd-based magnetic film properties, it is essential to know how $H_2$ is incorporated into the sample. Strong interactions of H atoms and its isotopes with neutrons make neutron reflectivity measurements a precise method to determine structural and magnetic changes that may take place inside the sample with depth resolution at the nanometer scale.[16] In contrast to traditional magnetometry and structural measurements, neutron reflectivity allows direct determination of where $H_2$ or $D_2$ is incorporated and how the magnetic profile in the sample is affected. It also allows for the determination of lattice expansion upon $H_2$ or $D_2$ absorption, thus helping to gauge the importance of magnetoelastic effects. Obtaining this information, however, requires a detailed quantitative analysis of the neutron reflectivity data. We note that x-ray scattering is not very sensitive to $H_2$, and therefore an indirect determination of $H_2$ absorption would normally rely on lattice expansion measurements if they occur.

Here we present PNR measurements in air or helium (He) and $D_2$ atmospheres on two Co/Pd multilayer samples with PMA. Each sample was measured with the magnetization vector forced to be either totally or partially in the plane of the sample by applying a magnetic field in the sample plane. X-ray reflectivity was used to verify the nuclear ordering structure. Magnetic PNR data were complemented by direct magnetization measurements in He and $H_2$ atmospheres obtained using standard magnetometry. Our results indicate that electronic effects resulting from $H_2$ or $D_2$ absorption are responsible for a decrease in the PMA and saturation magnetization of the samples.

## II. EXPERIMENTAL PROCEDURES

### A. Sample Growth

Sapphire (110) substrates were cleaned with methanol and subsequently annealed at 1400 °C for 3 hours. Atomic force microscopy (AFM) showed that the resulting surface consisted of atomically-smooth terracesseparated by atomic steps.[17] Each sample was grown by DC sputtering at a base pressure less than $6.7 \times 10^{-7}$ mbar. During growth, the substrates were rotated about their surface normal to promote uniform layer thickness. Sample A consisted of a 35 Å Pd buffer layer grown at 300° C followed by 40 periods of Pd (21 Å)/Co (2.5 Å) capped with 35 Å of Pd grown at 200° C. Sample B, grown entirely at 300° C, was composed of a 27 Å Pd buffer layer, followed by 40 periods of Pd(21 Å)/Co(2.5 Å) bilayers, and capped by a 27 Å Pd layer. Both samples were grown in an Ar partial pressure of 4.0 x $10^{-3}$ mbar. Nominal layer thickness values quoted above were determined from x-ray reflectivity (XRR) of ~200 Å thick pure Pd and Co calibration films. Because the roughness at the interfaces was comparable to the thin layer thicknesses of the periodic layers, the effective layer thicknesses and compositions were significantly different from the nominal values, thus precluding typical x-ray and neutron reflectivity structural determination. Therefore, an alternative method of analysis for these parameters was used (see Appendix A).

**B. X-ray Structural Characterization**

X-ray diffraction (XRD) using Cu $K_\alpha$ radiation (wavelength 1.5418 Å) was used to determine the crystal quality of the sample along the growth direction. XRR data analysis was used to obtain depth profiles of the scattering length densities (SLD) and thus deduce structural parameters (layer thicknesses and interfacial roughness) to compare with and validate PNR structural results, as outlined in Appendix A.

**C. Magnetization Measurements**

Magnetic moment measurements were performed using vibrating sample magnetometry (VSM) at room temperature in one atmosphere of He or $H_2$ with an external magnetic field, *H*, applied both in and out of the sample plane. With the sample in a He atmosphere and *H* applied in the sample plane, a SQUID magnetometer was used to measure magnetization hysteresis curves in fields up to 7 T to determine the saturation field of each sample. Magnetic force microscopy (MFM) images of the magnetic domains were obtained using a commercial scanning probe microscope at zero field after magnetizing the samples normal to the sample surface.

## D. Polarized Neutron Reflectivity Measurements

PNR experiments were performed on the Asterix reflectometer at the Los Alamos Neutron Science Center. The reflectometer views a partially coupled cold neutron moderator[17] through a $^{58}$Ni guide. The scattering angle in the horizontal plane $2\theta$ was measured using a one-dimensional position sensitive detector (20 cm long) located approximately 2.5 m from the sample. The neutron wavelength, ranging from 4 Å to 12 Å, was measured using a time-of-flight technique.[18] A super-mirror polarization cavity, which provides >96% degree of polarization, was used to control the incident neutron polarization. Corrections were made to take into account imperfections in the neutron beam polarization and wavelength variation of the neutron spectrum.[19]

Figure 1 shows the PNR scattering geometry. A magnetic field $\boldsymbol{H} = H\hat{x}$ provided the magnetization with a component in the plane of the sample and perpendicular to the scattering wavevector, $\boldsymbol{Q}$. This was necessary because the magnetic neutron scattering cross-section is in general only sensitive to components of $\boldsymbol{M} \perp \boldsymbol{Q}$.[18] The polarized neutron beam was incident on the sample at an angle $\alpha$ with the magnetic moment of incoming neutrons aligned parallel or anti-parallel to $H$.

For sample A, PNR measurements were performed in the presence of the polarization analyzer at fields of 6 T and 0.65 T in $H_2$ and $D_2$ atmospheres. Reflectivity cross-sections $R^{++}$ and $R^{--}$ were measured with the polarization vector of incident and reflected neutron beams parallel (+ +) or anti-parallel (- -) to the external magnetic field, respectively. As the superconducting magnet dewar configuration used for these measurements introduced a substantial amount of background noise, data capture was limited to a wavevector transfer of $Q \leq 0.15$ Å$^{-1}$. Spin-flip scattering ($R^{+-}$ and $R^{-+}$) measured close to the critical edge was at least two orders of magnitude smaller than $R^{++}$ and $R^{--}$ scattering at 0.65 T, showing that the magnetization of the layers did not have a significant component perpendicular to the field.

Sample B was enclosed in a displex cryostat and PNR measurements were carried out in a field of 0.65 T. An external field was produced by an electromagnet and the polarization analyzer was not used. Spin-flip scattering was assumed to be negligible, i.e. the component of the magnetization perpendicular to the external field was assumed to be small, as was observed for sample A.

For measurements obtained at $\mu_o H = 6$ T, the magnetic moment of the sample was confined to be within the plane of the film, whereas for $\mu_o H = 0.65$ T, the magnetic moment had only a component in the plane of the sample. All PNR measurements were performed at room temperature in a pressure of one atmosphere of air, He, or $D_2$.

$D_2$ was chosen instead of $H_2$ for the neutron reflectivity measurements because $D_2$ has a large positive scattering length ($\underline{b}_D = 6.671 \times 10^{-5}$ Å). This increases the contrast when compared to $H_2$, which has a smaller, negative scattering length ($\underline{b}_H = -3.7406 \times 10^{-5}$ Å).[20] In our model, the film was allowed to expand freely normal to the sample surface upon $D_2$ absorption. In-plane expansion was assumed to be negligible, as it was hindered by adhesion forces between the substrate and the multilayer.[16]

PNR data were fit to extract the depth profile of the projection of the magnetization along the polarization axis of the neutron beam and the nuclear depth profile before and after $D_2$ absorption. Because the layer thicknesses were small, obtaining sensible layer thickness and interface roughness parameters (i.e., interface roughness smaller than layer thickness) was not possible using standard methods. Therefore the data were analyzed and parameters obtained from the $\rho_n$ and $\rho_m$.[21] In addition, uncertainties of the fitting parameters were obtained by a Monte Carlo simulation procedure.[22] This methodology is described in Appendix A.

**III. RESULTS**

## A. Structure According to X-ray Diffraction and X-ray Reflectivity

X-ray diffraction showed highly oriented growth in the Pd (111) direction [Figure 2 (a) and (b)]. Multilayer periodicities obtained from the separation of the multilayer peaks in Figure 2(a) (23.2 Å) and Figure 2 (b) (23.7 Å) agreed well with those obtained from XRR and PNR data for sample A ($t_{Co\ stack} + t_{Pd\ stack} = 23.5 \pm 1.4$ Å) and sample B ($t_{Co\ stack} + t_{Pd\ stack} = 23.5 \pm 2.5$ Å). High angle XRD of sample A showed the presence of a Pd (200) phase, which was absent in the spectrum of sample B, although the presence of multilayer peaks were more prominent in sample A than in sample B.

XRR was used to determine the non-magnetic structure. Since a wide range of $Q$ values is accessible with XRR, it is possible to accurately deduce the nuclear structures of the samples. Layer nomenclature was defined as shown in Figure 3. The PdO layer at the sample/air interface accounted for oxidation after exposing the sample to air. In order to fit the XRR data for sample A, the thicknesses of the Pd 1, Pd S and Pd 2 layers were constrained to the same value [Figure 3(a)]. For sample B, the thicknesses of Pd 1, Pd S1, Pd S2 and Pd 2 were constrained to the same value, as were the values of Co 1, Co S1 ,Co S2 and Co 2 [Figure 3(b)].

## B. Magnetometry Measurements

Magnetization measurements for both samples are shown in Figure 4. By comparing the measurements with $H$ perpendicular and parallel to the sample plane, we found that 35.7% and 53.8% of the magnetization was in the plane of the sample at $\mu_0 H = 0.65$ T for samples A and B, respectively. Square loops measured with the field applied perpendicular to the sample plane confirm the presence of a large out-of-plane anisotropy.[23] SQUID magnetometry revealed that the in-plane saturation field of sample A was 5.5 T (not shown). The observed decrease of the magnetization in sample B as the field decreased from saturation (for $H > 0$ and $H \perp$ to the sample surface) was due to the formation of magnetic domains.[14] This was verified by MFM images of sample B (Figure 5), which indicated the presence of irregular striped domains characteristic of ferromagnetic Co/Pd multilayers with PMA.[24] Sample A displayed a larger remanence and coercivity than sample B, possibly as a result of greater atomic intermixing at the interfaces (deduced from XRR and PNR results discussed below), which is known to result in more pinning centers that obstruct domain growth and propagation.[25] As a result, sample A showed no domain structure via MFM on the scale examined.

Magnetization measurements showed that when $H$ was applied in the plane of the samples, there was a net increase in magnetization component along $H$ upon $H_2$ absorption at $\mu_0 H = 0.65$ T, the increase being larger in sample A. With $H \perp$ to the sample surface, the saturation magnetization decreased in both samples, although the effect was greater for sample A. Therefore, the increase in the unsaturated state, with $H$ in the plane of the samples, must have been due to a decrease in the PMA.

SQUID and VSM magnetization measurements thus provide clear evidence for global changes in the magnetic properties of both samples upon $H_2$ absorption; but it is difficult to determine where the $H_2$ absorption is most prevalent, which layers are affected, or whether the Pd layers are magnetized. Moreover, it is not possible to determine whether magnetoelastic effects or direct modification of the density of states at the Fermi level are responsible for these changes.

### C. PNR Results

In the PNR data analysis, parameters were constrained so that $\sigma_{\text{buffer}}=\sigma_{\text{Pd 2}}$ and $t_{\text{Pd }1} = t_{\text{Pd }2} = t_{\text{Pd S}}$. Layers Pd 1 and Pd 2 were introduced to account for possible proximity magnetic effects of adjacent Co layers on the Pd buffer and top capping layers, respectively.[25, 26] An interface roughness was also introduced to separate the non-magnetic Pd buffer layer from magnetic Pd 1 layer and the magnetic Pd 2 layer from Pd top layer. This interface was due to a purely magnetic contrast. In sample B, a slightly different model was used because$R^-$ data exhibited a half-order Bragg peak. This means that the magnetization of the Co/Pd stack structure varied with a periodicity twice that of the nuclear multilayer PNR component. Consequently, alternating Co/Pd stack layers were fit with independent magnetizations. The magnetization of layers Co1 and Co2 were fit independently from those of the Co/Pd stack, which were surrounded by the thicker Pd buffer and top layers. The magnetic scattering lengths of the Pd 1 and Pd 2 layers were fit independently as was done for sample A.

PNR data obtained from sample A with the fit to the model are shown in Figure 6 (0.65 T) and Figure 7 (6 T). For clarity, neutron reflectivity data are shown as a product with $Q^4$, which compensates for the well-known power law decrease in reflectivity with increasing $Q$.[17] Qualitatively, the decrease in the period of oscillations in the low-$Q$ regime upon $D_2$ absorption indicates an increase in the total thickness of the sample. The same observation can be made for sample B. Figures 8 and 9 show the high and low $Q$ portions of the data and fits from sample B, respectively. The fact that the position of the multilayer peak at $Q$ = 0.27 Å$^{-1}$ remained unchanged indicates that the Co/Pd multilayer period did not change upon $D_2$ absorption. Therefore, the increase in total sample thickness is solely due to an expansion of the Pd top and buffer layers.

Nuclear SLD profiles $\rho_n$ and their derivatives obtained from fits of the PNR data are shown in Figure 10 for sample B. A Similar $\rho_n$ profile was obtained for sample A, but analysis of these data was less reliable due to the absence of the multilayer Bragg reflection because of our inability to measure at high $Q$ as discussed above. Positions of the interfaces, determined from the locations of the maxima and minima in $d\rho_n/dz$, are indicated by vertical dotted lines.

Table I and Table II summarize thicknesses, interface roughness, and scattering length parameters of each layer determined from $\rho_n$ profiles and their derivatives. Note that PNR and XRR measurements in air and He yielded parameters which agreed with each other to within their respective uncertainties.

Structural parameters obtained for sample A at 6 T and 0.65 T also agree to within the uncertainties shown in Table I.

Values of $\rho_n$ were used to determine the stoichiometry of each layer independently from Equation 3. The Pd buffer and top layers' SLDs correspond to bulk Pd in both samples A and B. For reference, we note that the accepted values of bulk Pd for neutron and x-rays are $4.01 \times 10^{-6}$ Å$^{-2}$ and $87.9 \times 10^{-6}$ Å$^{-2}$, respectively.[18, 27] Therefore, in sample A, the center of the Pd stack layers consisted of 95% Pd and the Co stack layers were 12% Co. Sample B consisted of Pd stack layers with 89% Pd and Co stack layers with 30% Co.

Upon $D_2$ absorption, there was a statistically significant increase in the thickness of the Pd buffer and top layers in both samples. Results also indicate statistically insignificant changes in the Co and Pd stack layer thicknesses. A noteworthy decrease of the PdO layer thickness occurred in both samples, which can be attributed to reduction by deuterium.[28] In the $d\rho_n(z)/dz$ profile of sample A (Figure 10), the position of the PdO peak disappears completely, while for sample B there was still a peak, indicating that the PdO was not completely reduced.

Comparing $\rho_n$ and thickness change before and after $D_2$ absorption (Figure 11), the ratio of the number of deuterium atoms to Pd atoms, $C_D$, was estimated in each layer using Equation A4. $C_D$ was found to be 0.53 and 0.30 for sample A and 0.75 and 0.52 for sample B in the Pd buffer and Pd top layers, respectively, confirming that there was significant $D_2$ absorption in these layers. The value of $C_D$ for the Pd and Co stacks in sample A and B was approximately $0.02 \pm 0.005$. The relatively small value of $C_D$ for the Pd and Co stack explains the lack of significant lattice expansion and implies that D was probably absorbed into interstitial sites while the film remained in the α-phase, where lattice expansion is minimal.[4, 29] This might be due to the presence of Co in the Pd stack layers which could have decreased the heat of deuterium absorption with respect to the Pd top and Pd bottom layers.[30, 31]

The magnetic SLD profile, $\rho_m$, for samples A and B are shown in and Figure 12 and Figure 13, respectively. The maxima and minima in $\rho_m$ correspond to Co and Pd layers, respectively. Table III and Table IV summarize the magnetization in the Co and Pd layers for sample A at 6 T and 0.65 T. Magnetizations in the Co and Pd layers of sample B at 0.65 T are summarized in Table V. Uncertainties in the magnetization of the layers in sample A were significantly larger than those in sample B, again due to the limited data collection range and lack of a Bragg reflection. Magnetizations of the stack layers had overlapping error bars, making it difficult to determine which layer's magnetization changed significantly upon $D_2$ absorption. At the saturation field of 6 T, Co layers had a magnetization lower than that of bulk Co ($1.44 \times 10^3$ kA/m) because of dilution with Pd. Interestingly, the Pd stack layers in sample A at 6 T (saturation field) have lower magnetization than the Pd stack layers of sample B at 0.65 T, probably as a result of the higher purity of the nominal Co layers in sample B, causing a stronger proximity effect on the Pd. At 0.65 T, both sample A and sample B had lower in-plane magnetization, in agreement with VSM measurements.

As shown in Figure 13 and Table V, most of the modulation in the magnetization of sample B occurred due to different values of the Pd S1 and Pd S2 layers (111 kA/m and 167 kA/m, respectively). These layers were also the most affected by $D_2$ absorption, increasing by nearly 16% in both cases, corresponding to at least one standard deviation for Pd S1 and Pd S2. On the other hand, the Co S1 and Co S2 layer magnetizations remained approximately constant. We also note that $M$(Co1) = $M$(Co S2) and $M$(Co2) = $M$(Co S1) to well within the uncertainty of the measurements. Therefore we conclude that the magnetizations of the Co layers at the bottom and top of the sample were the same as those of the stack.

Doubling of the magnetic period of the multilayer can be understood in terms of a modulation of the PMA within the stack. Since the magnetization was not saturated, layers with weaker anisotropy tilted more strongly towards the field direction, causing a $Q_{1/2}$ peak to appear. An inter-layer magnetic interaction, which modulated the PMA generated by the Co/Pd interface, could have been responsible, but determination of the origin of this effect requires a more thorough study. Our results also indicate that the modulation grew stronger with $D_2$ absorption. This is evident in Figure 8, which shows that the $Q_{1/2}$ peak became more pronounced, and in the magnetic SLD profile in Figure 13, which shows the increase in magnetic contrast between adjacent minima, corresponding to Pd layers.

Total sample magnetization variation upon $D_2$ absorption was further verified by integrating the magnetic SLD profile and comparing the resulting moment with the moment obtained via VSM measurements. In order to obtain an accurate magnetization measurement, VSM data were averaged over several minutes in He and $H_2$ atmospheres. Results for sample A are shown in Table VI. VSM data obtained with $H$ normal to the sample surface had a total magnetization aligned with the applied field. Comparing this value with the 6 T PNR measurement with $H$ in the sample plane, where $M$ is also saturated, revealed that $M$ decreased upon $H_2/D_2$ absorption, while the reverse was true when the sample was not saturated. The quantitative results for PNR and VSM agreed to within their uncertainties for the unsaturated measurements and also revealed that the change was slightly larger for PNR at saturation. A decrease in the total magnetization upon $D_2$ absorption at saturation in sample A was also determined from the PNR data, as seen in the magnetic measurements, but the percentage increase obtained by PNR was larger than that seen in VSM. Table VII shows the results for sample B, measured with the field in the plane with $M$ not at saturation. The PNR and VSM measurements both show an increase in $M$ when $H_2/D_2$ is absorbed, in agreement with the results from sample A.

## IV. DISCUSSION

To summarize our most important experimental results, we have found that: 1) $D_2$ absorption occurred throughout both samples; 2) the multilayer stack absorbed $D_2$ but did not expand along the growth direction; 3) both the Pd and Co layers were magnetized and their in-plane magnetization increased when exposed to $H_2$/$D_2$ at low applied magnetic fields but decreased at saturation.

PNR data indicated that the Pd layers in the multilayer stack were ferromagnetic. It is well known that Pd is paramagnetic with high magnetic susceptibility, i.e. it is on the border of being ferromagnetic and can undergo spontaneous spin polarization when in proximity to ferromagnetic materials. In particular, it has been shown previously that there is a giant magnetic enhancement of Pd of up to 0.4 $\mu_B$ in Pd/Fe thin films.[32, 33] In Co/Pd multilayers, it is known that Pd atoms become polarized in the vicinity of Co atoms, resulting in the magnetization of the Pd layers.[25, 26]

Our measurements show that the in-plane magnetization increased upon $D_2$ absorption in both samples, but the opposite effect was observed in the out-of-plane magnetization, where the magnetization saturated at approximately 0.1 T and 0.4 T for sample A [Figure 4(a)] and sample B [Figure 4(b)], respectively, due to the PMA of the sample. The increase in *M* at 0.65 T for the in-plane VSM measurements must therefore have been due to a change in the magnetic anisotropy of the system, which is consistent with a decrease in the PMA. Similar increases in *M* were observed in both samples at 0.65 T upon deuterium absorption (Table VI and VII). VSM and PNR measurements were in agreement with each other to within their uncertainties.

The change in absolute magnetization at saturation observed with *H* applied perpendicular to the surface is consistent with the previously observed decrease in magnetic susceptibility in Pd upon $H_2$ absorption,[34] which has been interpreted by Mott as filling the *d*-electron holes with electrons donated by absorbed $H_2$.[35] Another possibility is given by Mydosh[36] who found that in Fe/Pd alloys the long-range RKKY coupling between Fe atoms is significantly reduced with $H_2$ absorption. If this were the case, the RKKY interaction within the Co layers, where interdiffusion is significant due to their small thickness, must contribute to the overall magnetization.

Regarding the effects of $H_2$ or $D_2$ absorption on the PMA, prior work in Co/Pd multilayers has shown that it is highly dependent on the interface structure,[37] with magnetostrictive effects induced by interfacial strain playing a key role reported in one instance[38] and the existence of an interface itself, however diffuse, in another.[14] The fact that there was no measurable expansion of the Co/Pd stack upon $D_2$ absorption implied that magnetostrictive effects were small in our samples. Thus, the changes in the PMA must be a result of the interface structure with electron transfer from the absorbed deuterium to the Co/Pd multilayer. This conclusion is in agreement with work in other metallic multilayers that absorb $H_2$ where changes in their magnetic properties are also believed to result from electron transfer rather than from magnetoelastic effects.[39]

The change in magnetization was larger in sample A than in sample B. One possible reason is that the larger Pd concentration in the Co layers in the stack of sample A increased the amount of $H_2$ absorption, as indicated in Figure 11, thus enhancing the hydrogen-induced magnetization change. One cannot discount, however, the possibility that this may also be due to the slightly thicker Pd top and Pd buffer layers in sample A, which could also increase the $H_2$ uptake.

Finally, we note that magnetization measurements in the VSM as the $H_2$ was cycled in and out of similar samples also revealed that the change in magnetic moment upon $H_2$ absorption and desorption was completely reversible. These data will be presented in a future publication specifically dealing with this subject.

## V. CONCLUSIONS

Scattering length densities obtained from PNR measurements were analyzed to determine structural parameters and depth dependence of the magnetization in a Co/Pd multilayer in order to understand the effects of $H_2/D_2$ absorption on two different samples. Results from the PNR fits indicated an increase in the total thickness of both samples. Most of the increase in thickness occurred at the buffer and top Pd layers, however, and yet the $\rho_n$ depth profile indicated that deuterium absorption occurred throughout the sample. The magnetic SLD showed a modulation of the magnetization with a period equal to twice the Co/Pd bilayer thickness at a field of 0.65 T in sample A. PNR measurements and the magnetization measurements confirmed an increase in the in-plane component of $M$ when the samples were exposed to $D_2$ or $H_2$ in an in-plane field of 0.65 T. Magnetization measurements and PNR at saturation showed that the saturation magnetization decreases with $H_2$ and $D_2$ absorption. These results indicate that $H_2$ or $D_2$ absorption in Co/Pd multilayers causes changes in the electronic structure which results in lower PMA and total magnetization.

## ACKNOWLEDGEMENTS

This work was supported by the DOE grant DE-PS02-07ER087-15 and the WVNano Initiative at WVU. The Los Alamos Neutron Science Center facility at the Los Alamos National Laboratory is funded by the US Department of Energy Office of Basic Energy Sciences. Los Alamos National Laboratory is operated by Los Alamos National Security LLC under DOE Contract DE-AC52-06NA25396. The authors thank Matts Björck for his valuable suggestions in using his program, GenX, for fitting the neutron data.

## APPENDIX A: XRR AND PNR DATA ANALYSIS PROCEDURES

Neutron reflectivity probes variations in the neutron SLD as a function of depth of the structure (the $z$ direction parallel to the surface normal). The SLD, $\rho\ (z)$, and its nuclear and magnetic components $\rho_n(z)$ and $\rho_m(z)$, are given by

$$\rho_\pm(z) = \rho_n(z) \pm \rho_m(z), \tag{A1}$$

$$\rho_n(z) = \textstyle\sum_i^s N_i(z) b_i(z), \tag{A2a}$$

$$\rho_m = c \textstyle\sum_i^s N_i(z) \mu_i(z) \tag{A2b}$$

where $s$ is the number of distinct isotopes, $N_i$, $b_i$ and $\mu_i$ are the number density, scattering length, and magnetic moment of the $i$-th species, respectively, and $c$ = $2.645 \times 10^{-5}$ Å/$\mu_B$.[18] For PNR, the reflectivities measured with the incoming neutron spins parallel and antiparallel to the applied field when scattered ($R^+ = R^{++} + R^{+-}$ and $R^- = R^{--} + R^{-+}$, respectively) yield $\rho_m(z)$ with the positive and negative signs in Eqn. 1.[18] By simultaneously fitting the $R^+$ and $R^-$ data, the $\rho_+$ and $\rho_-$ SLD's are generated, and the nuclear and magnetic profiles can be extracted from $\rho_n = (\rho_+ + \rho_-)/2$ and $\rho_m = (\rho_+ - \rho_-)/2$. From $\rho_n$ obtained from PNR and XRR data, the actual stoichiometry of the Co and Pd stacks due to interface diffusion can be deduced from

$$\rho_n^{neutron}(z) = N_{Co}(z)b_{Co} + N_{Pd}(z)b_{Pd}, \quad \text{(A3a)}$$

$$\rho_n^{XRR}(z) = N_{Co}(z)r_0Z_{Co} + N_{Pd}(z)r_0Z_{Pd}, \quad \text{(A3b)}$$

where $r_0$is the classical radius of an electron (2.8 fm) and $Z$ is the atomic number.[40] Since the number density $N_i$(z) is the same for XRR and PNR, the value obtained from $\rho_n$can be used to estimate the compositions of the layers and interfaces independently.

The concentration of deuterium $C_D$ (number of deuterium atoms/ number of Pd atoms) can be calculated from[16]

$$c_D = \left[\frac{\rho_{n(Pd+D)}}{\rho_{n(Pd)}}\frac{t_{Pd+D}}{t_{Pd}} - 1\right]\frac{b_{Pd}}{b_D}, \quad \text{(A4)}$$

where $t_{Pd}$ and $t_{Pd+D}$ are the thickness values of the Pd layer in the pristine and loaded states, respectively.

Neutron reflectivity data were fitted using GenX,[41] a software package which uses the Parratt recursion formalism[42] for simulation and a genetic algorithm for parameter optimization. Fitted parameters were obtained using the minimization of chi-squared, $\chi^2$, defined in the traditional way as $\chi^2 = \sum_{i=0}^{N-1}(y_i - \mu_i)^2/s_i^2$, where $N$ is the number of data points, $\mu_i$ the $i$-th data point generated by the model, $y_i$ the $i$-th measured data point, and $s_i$ the uncertainty for each data point, the latter being the square root of the number of counts.

Because the thickness of the layers was comparable to the interface roughness, our determination of the structure was based on analyzing the SLD profile of the sample obtained from XRR and PNR data. In this approach, the SLD profile was first generated by fitting the effective SLD of the layers, the thickness of the layers, and interface roughness parameters using Parratt's formalism. The actual layer thickness and roughness parameters were obtained from $d\rho_n/dz$.[21] The layer thickness was defined as the distance between maxima and minima in $d\rho_n/dz$. The interface roughness, defined as the effective width of the interface, was defined as the square-root of the variance calculated from the probability distribution represented by $d\rho_n/dz$. Explicitly, this corresponds to

$$\sigma_i{}^2 = < z_i >^2 - < z_i{}^2 >, \tag{A5}$$

where $\sigma_i$ is the interface roughness parameter, and $< z_i >$ the position of the $i^{th}$ interface. Here $< z_i >$ was calculated as

$$< z_i > = \int_{z_{i1}}^{z_{i2}} \frac{d\rho}{dz} z dz / \int_{z_{i1}}^{z_{i2}} \frac{d\rho}{dz} dz, \tag{A6}$$

with a similar expression for $< z^2 >$. The integrals in Eqn. A6 were calculated numerically with integration limits $z_{i1}$ and $z_{i2}$ chosen to be the values of $z$ where $d\rho_n/dz$ crossed zero with a peak or trough in between them. The effective thickness of the $i^{th}$ layer was calculated using

$$t_i = < z_{i+1} > - < z_i >. \tag{A7}$$

Values of the SLD for the $i^{th}$ layer, $\rho_i$, were determined from the value of the SLD profile at the center of each layer, which was defined as

$$d_i = (< z_{i+1} > + < z_i >)/2. \tag{A8}$$

Uncertainty values for $\sigma_i$, $t_i$, and $\rho_i$ were determined by generating ten artificial sets of data with the same number of hypothetical data points as the measured data. These data sets were generated using a Monte Carlo simulation procedure consisting of a normal-distribution random number generator such that the data points tended to be within the measured error bars. The artificial data were fitted using the same procedure as the measured data (i.e., by analyzing the SLD profile) and ten values of $\sigma_i$, $t_i$, and $\rho_i$ were produced. The standard deviation of these values gave the uncertainty for each fitted parameter.[22] A similar procedure, also using GenX, was used to determine the uncertainties of layer thickness and interface roughness parameters obtained from XRR.

## Tables

**Table I** Results of fitting polarized neutron reflectivity data measured in 1 atm of He and deuterium, as well as x-ray reflectivity measured in air for sample A.

| **Parameter** | **PNR Air** | **PNR $D_2$** | **XRR Air** |
|---|---|---|---|
| $\sigma_{sub}$ (Å) | 1.8 ± 1.2 | 4.6± 2.1 | 2.6± 1.1 |
| $t_{Pd\ buffer}$ (Å) | 32.7± 2.7 | 50.0± 2.1 | 35.2± 1.4 |
| $\sigma_{Pd\ 1}$ (Å) | 3.4±1.8 | 4.2±1.5 | 5.2±1.0 |
| $\sigma_{Pd\ buffer}$ (Å) | 1.5±1.1 | 1.9±1.5 | - |
| $t_{Co\ stack}$ (Å) | 9.0± 2.1 | 7.5±1.3 | 9.5± 1.0 |
| $\sigma_{Co}$ (Å) | 3.6± 1.4 | 1.6 ± 1.4 | 3.4± 0.6 |
| $t_{Pd\ stack}$ (Å) | 14.5± 1.3 | 17.1± 1.6 | 14.0±1.0 |
| $\sigma_{Pd}$ (Å) | 2.1 ±1.1 | 2.2± 1.2 | 2.2±0.5 |
| $t_{Pd\ top}$ (Å) | 33.5± 2.2 | 45.0± 1.3 | 36.5± 1.0 |
| $\sigma_{Pd\ top}$ (Å) | 3.6± 1.4 | 5.3± 2.1 | 2.7± 0.7 |
| $t_{PdO}$ (Å) | 12.0 ± 2.0 | 4.0±1.1 | 12.0± 1.3 |
| $\sigma_{PdO}$ (Å) | 2.1± 1.5 | 2.2±1.3 | 5.4± 1.1 |
| $\rho_{n\ Pd\ top}$ ($10^{-6}$ Å$^{-2}$) | 4.01± 0.03 | 4.01± 0.02 | 88.1± 2.5 |
| $\rho_{n\ Co\ stack}$ ($10^{-6}$ Å$^{-2}$) | 3.85± 0.03 | 3.86 ± 0.02 | 82.8± 2.7 |
| $\rho_{n\ Pd\ stack}$ ($10^{-6}$ Å$^{-2}$) | 3.98± 0.03 | 4.07± 0.03 | 84.7± 1.9 |
| $\rho_{n\ Pd\ buffer}$ ($10^{-6}$ Å$^{-2}$) | 4.04± 0.02 | 4.21± 0.03 | 88.1± 2.2 |

**Table II** Results of fitting polarized neutron reflectometry data measured in 1 atm of air and deuterium, as well as x-ray reflectivity measured in air for sample B.

| Parameter | PNR Air | PNR $D_2$ | XRR Air |
|---|---|---|---|
| $\sigma_{sub}$ (Å) | 2.3 ± 1.9 | 3.2 ± 2.3 | 1.9 ± 1.1 |
| $t_{Pd\ buffer}$ (Å) | 19.9 ± 2.8 | 36.7 ± 2.8 | 22.4 ± 2.1 |
| $\sigma_{Pd\ 1}$ (Å) | 1.8±1.5 | 5.2 ± 2.1 | 2.3±0.8 |
| $\sigma_{Pd\ buffer}$ (Å) | - | - | - |
| $t_{Co\ stack}$ (Å) | 7.5± 1.9 | 11.4 ± 2.1 | 9.0 ± 1.5 |
| $\sigma_{Co}$ (Å) | 3.5 ± 1.1 | 2.4 ± 1.3 | 3.5 ± 0.5 |
| $t_{Pd\ stack}$ (Å) | 15.8 ± 1.8 | 12.6 ± 2.1 | 14.1± 1.6 |
| $\sigma_{Pd}$ (Å) | 1.7 ± 1.1 | 2.7 ± 1.0 | 1.9 ± 0.5 |
| $t_{Pd\ top}$ (Å) | 21.5 ± 4.1 | 34.1 ± 3.3 | 18.9 ± 2.1 |
| $\sigma_{Pd\ top}$ (Å) | 2.8 ± 1.6 | 7.3 ± 2.5 | 2.9 ± 0.7 |
| $t_{PdO}$ (Å) | 11.5 ± 2.2 | - | 9.5 ± 1.5 |
| $\sigma_{PdO}$ (Å) | 3.8 ± 1.4 | - | 5.2 ± 1.1 |
| $\rho_{Pd\ top}$ ($10^{-6}$ Å$^{-2}$) | 4.02 ± 0.04 | 4.02 ± 0.04 | 88.2± 2.2 |
| $\rho_{Co\ stack}$ ($10^{-6}$ Å$^{-2}$) | 3.48 ± 0.03 | 3.62 ± 0.03 | 80.0 ± 2.4 |
| $\rho_{Pd\ stack}$ ($10^{-6}$ Å$^{-2}$) | 3.81 ± 0.03 | 3.90 ± 0.02 | 84.9 ± 3.2 |
| $\rho_{Pd\ buffer}$ ($10^{-6}$ Å$^{-2}$) | 4.02 ± 0.03 | 4.21± 0.02 | 88.2 ± 2.9 |

**Table III** Results of fitting PNR data measured in 1 atm of helium and deuterium for sample A at 6 T. The values of $\boldsymbol{\rho_m}$ for each layer have been converted to units of magnetization.

| Layer | *M* PNR Air (kA/m) | PNR $D_2$ (kA/m) |
|---|---|---|
| Pd 1 | 70± 66 | 46± 35 |
| Pd S1 | 89± 68 | 45± 36 |
| Co S1 | 571± 99 | 701± 105 |
| Pd 2 | 68± 54 | 43± 38 |

**Table IV** Results of fitting PNR data measured in 1 atm of helium and deuterium for sample A at 0.65 T. The values of $\boldsymbol{\rho_m}$ for each layer have been converted to units of magnetization.

| **Layer** | ***M* PNR Air (kA/m)** | ***M* PNR D$_2$ (kA/m)** |
|---|---|---|
| Pd 1 | 46± 28 | 54± 38 |
| Pd S1 | 50± 35 | 52 ± 45 |
| Co S1 | 162 ± 94 | 252± 90 |
| Pd2 | 56± 33 | 51± 30 |

**Table V** Results of fitting PNR data measured in 1 atm of air and deuterium for sample B at 0.65 T. The values of $\boldsymbol{\rho_m}$ for each layer have been converted to units of magnetization.

| **Layer** | ***M* PNR Air (kA/m)** | ***M* PNR D$_2$ (kA/m)** |
|---|---|---|
| Pd 1 | 91 ± 7 | 106 ± 9 |
| Co1 | 208± 8 | 184 ± 10 |
| Pd S1 | 111 ± 8 | 129 ± 6 |
| Co S1 | 207 ± 7 | 212 ± 9 |
| Pd S2 | 167 ± 8 | 193 ± 6 |
| Co S2 | 213± 9 | 198 ± 12 |
| Co2 | 196 ± 10 | 193 ± 8 |
| Pd2 | 123 ± 12 | 105± 9 |

**Table VI** Comparison of total magnetic moment in helium and deuterium measured with the field applied in the plane, as determined by PNR and VSM measurements for sample A.

| Field | Measurement | $m_{He}$ ($10^{-7}$ A m$^2$) | $m_{D2}$ ($10^{-7}$ A m$^2$) |
|---|---|---|---|
| 0.65 T in plane | PNR | 4.9±0.4 | 5.8±0.5 |
| 0.65 T in plane | VSM | 5.29±0.01 | 6.00±0.01 |
| 6 T in plane | PNR | 15.0±0.5 | 13.1±0.4 |
| 0.65 T out of plane | VSM | 14.90±0.01 | 14.39±0.01 |

**Table VII** Comparison of total magnetic moment in air and deuterium measured at 0.65 T with the field applied in the plane, as determined by PNR and VSM measurements for sample B.

| Measurement | $m_{air}$ ($10^{-7}$ A m$^2$) | $m_{D2}$ ($10^{-7}$ A m$^2$) |
|---|---|---|
| PNR | 6.9 ± 0.2 | 7.5 ± 0.2 |
| VSM | 7.22 ± 0.01 | 7.37 ± 0.01 |

## Figures

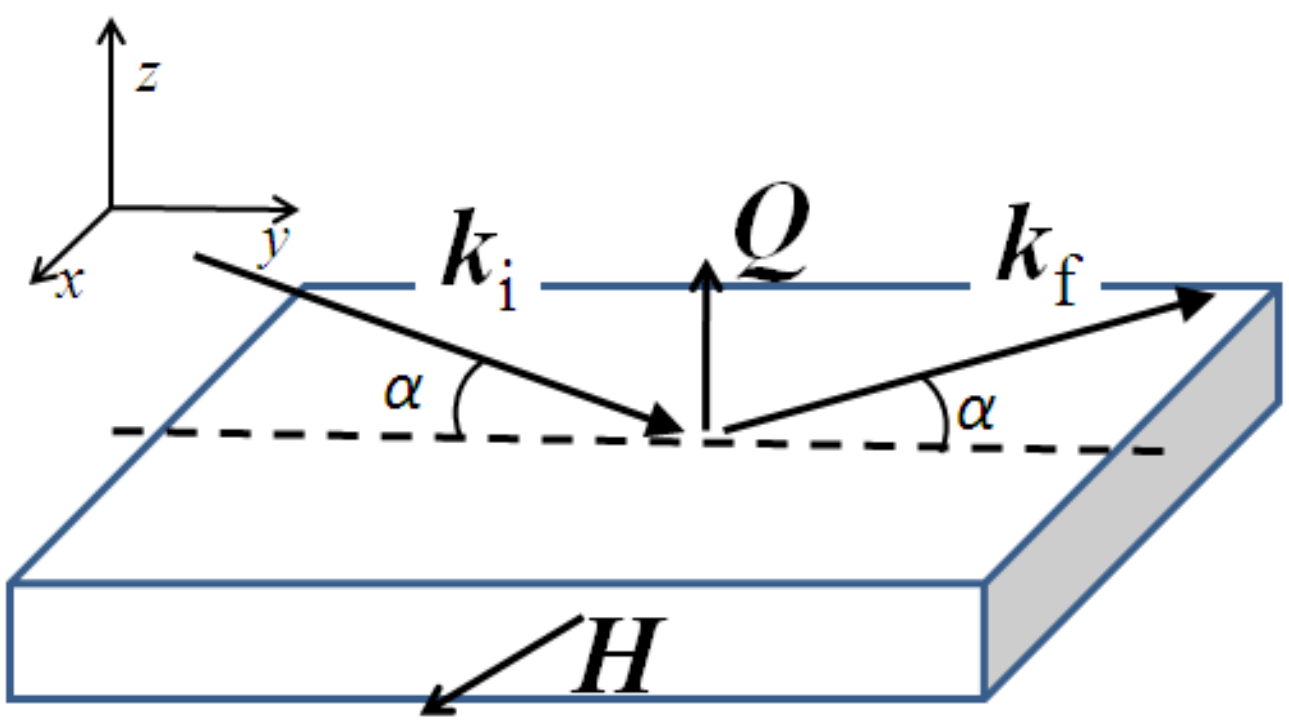

**Figure 1** (Color online) Diagram showing the scattering geometry for the PNR experiment. The magnetic field $\boldsymbol{H}$ is applied in the plane of the sample along the $x$-axis. For specular reflectivity, the angle of incidence of the neutrons is identical to the angle of reflection $\boldsymbol{\alpha}$. The scattering wavevector transfer $\boldsymbol{Q} = \boldsymbol{k_f} - \boldsymbol{k_i}$ is parallel to the $z$-axis and perpendicular to the sample surface. Since magnetic neutron scattering is sensitive to the components of $\boldsymbol{M}$ perpendicular to $\boldsymbol{Q}$, only the components of $\boldsymbol{M}$ in the plane of the sample ($x$-$y$ plane) are probed by PNR.

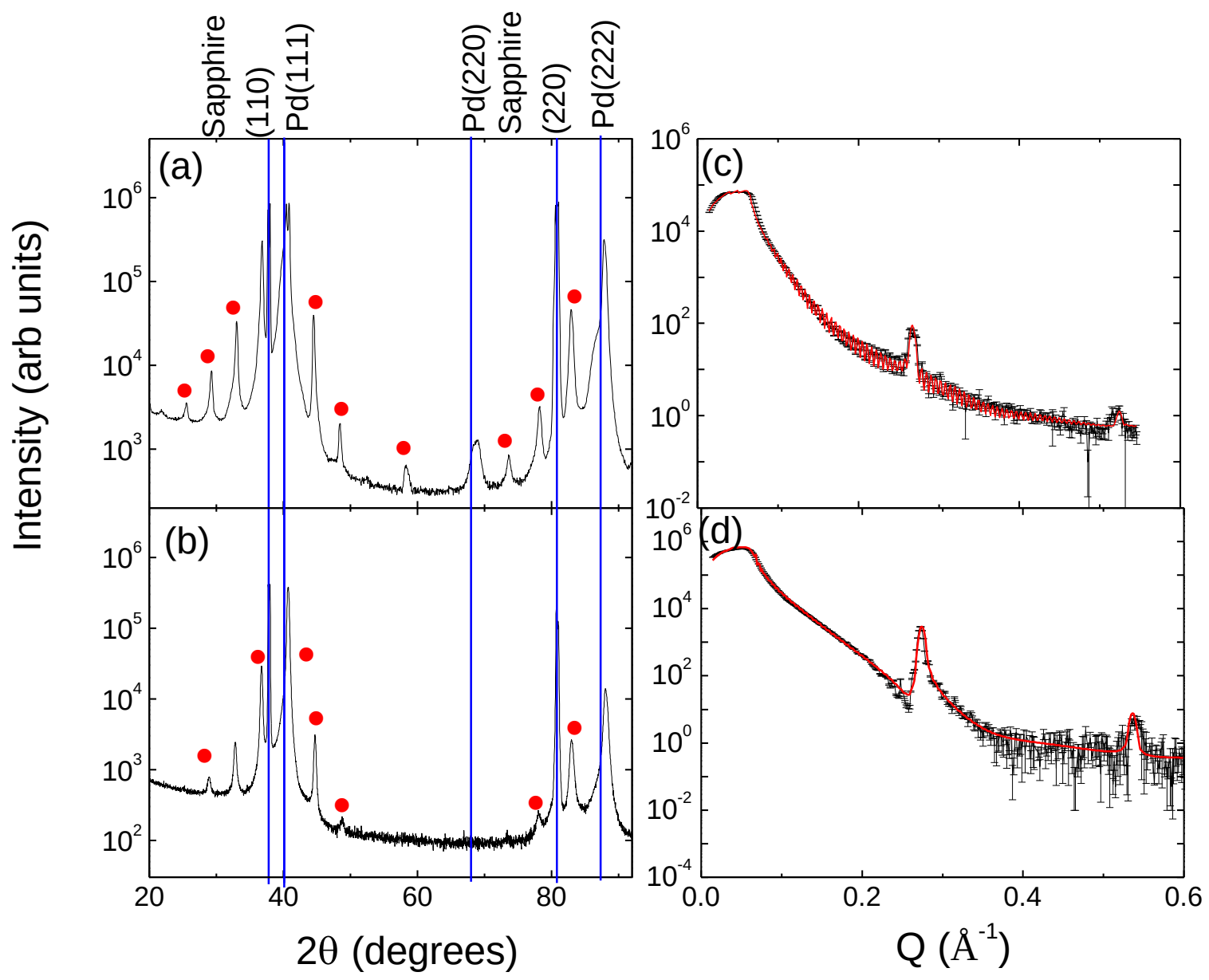

**Figure 2** (Color online) (a) and (b): High angle x-ray diffraction of the Co/Pd multilayer for samples A and B, respectively. The expected positions of the sapphire substrate peaks and Pd bulk buffer layer peaks are indicated. Multilayer peaks are indicated by a red dot. (c) and (d): X-ray reflectivity measurements of the sample A and B respectively. The black dots in the reflectivity graphs are the data and the red lines are the fit to the data.

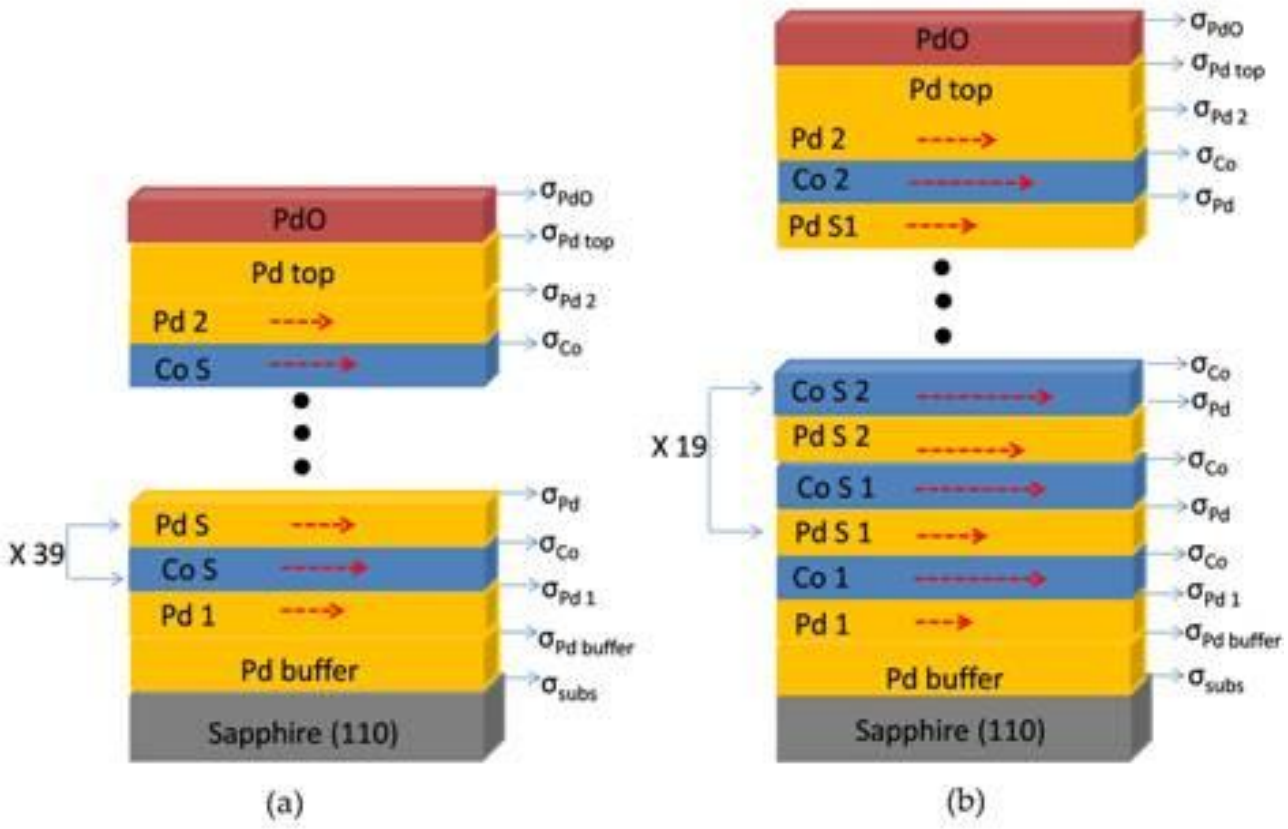

**Figure 3** (Color online) Sketch of (a) sample A and (b) sample B used in XRR and PNR models. The location of the interface roughness parameters $\sigma$ and the layers used as fitting parameters are illustrated. The dashed red arrows indicate the magnetization used in the PNR model only.

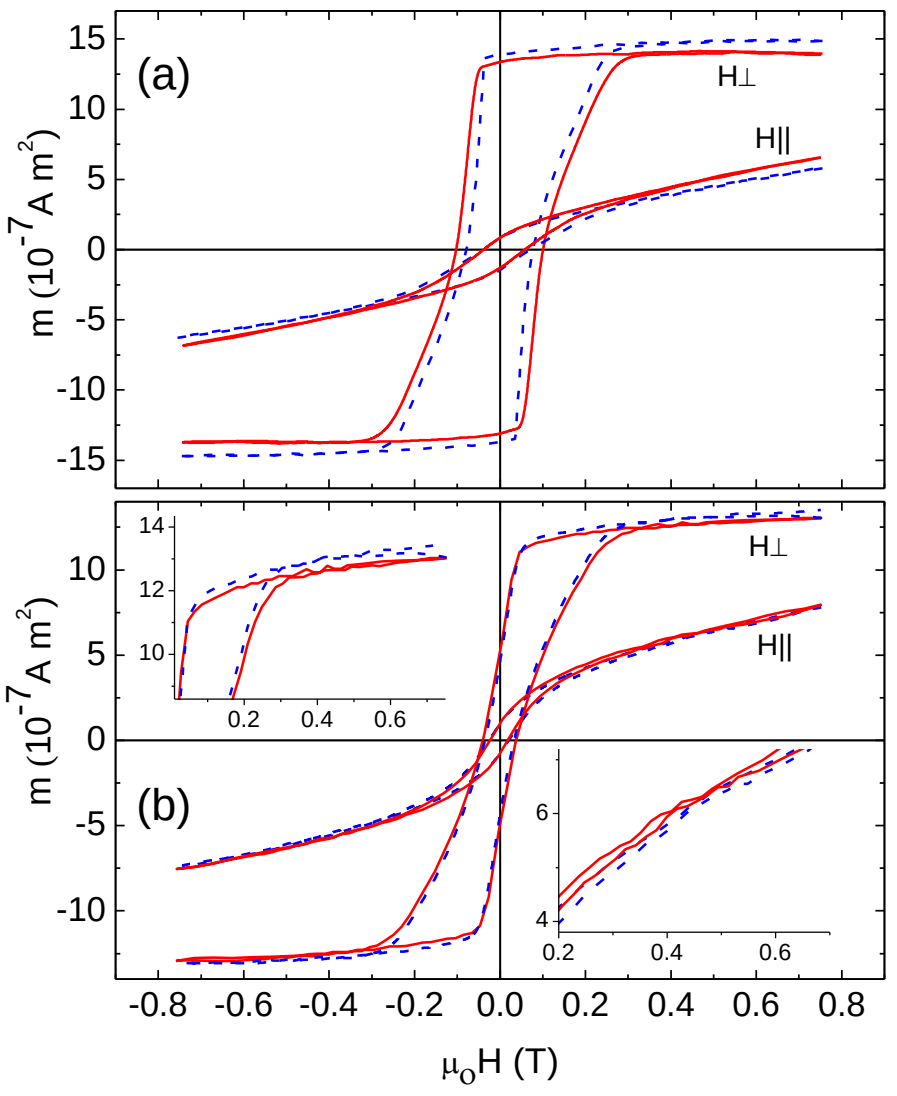

**Figure 4** (Color online) Magnetic moment measurements in 1 atm of He (blue dashed curves) and $H_2$ (red solid curves) with the magnetic field applied perpendicular (H⊥) and parallel (H||) to the sample surface. (a) Data for sample A, (b) data for sample B. Top left and bottom right insets in (b) are close-up views of the data in (b) for the H⊥ and H|| configurations,

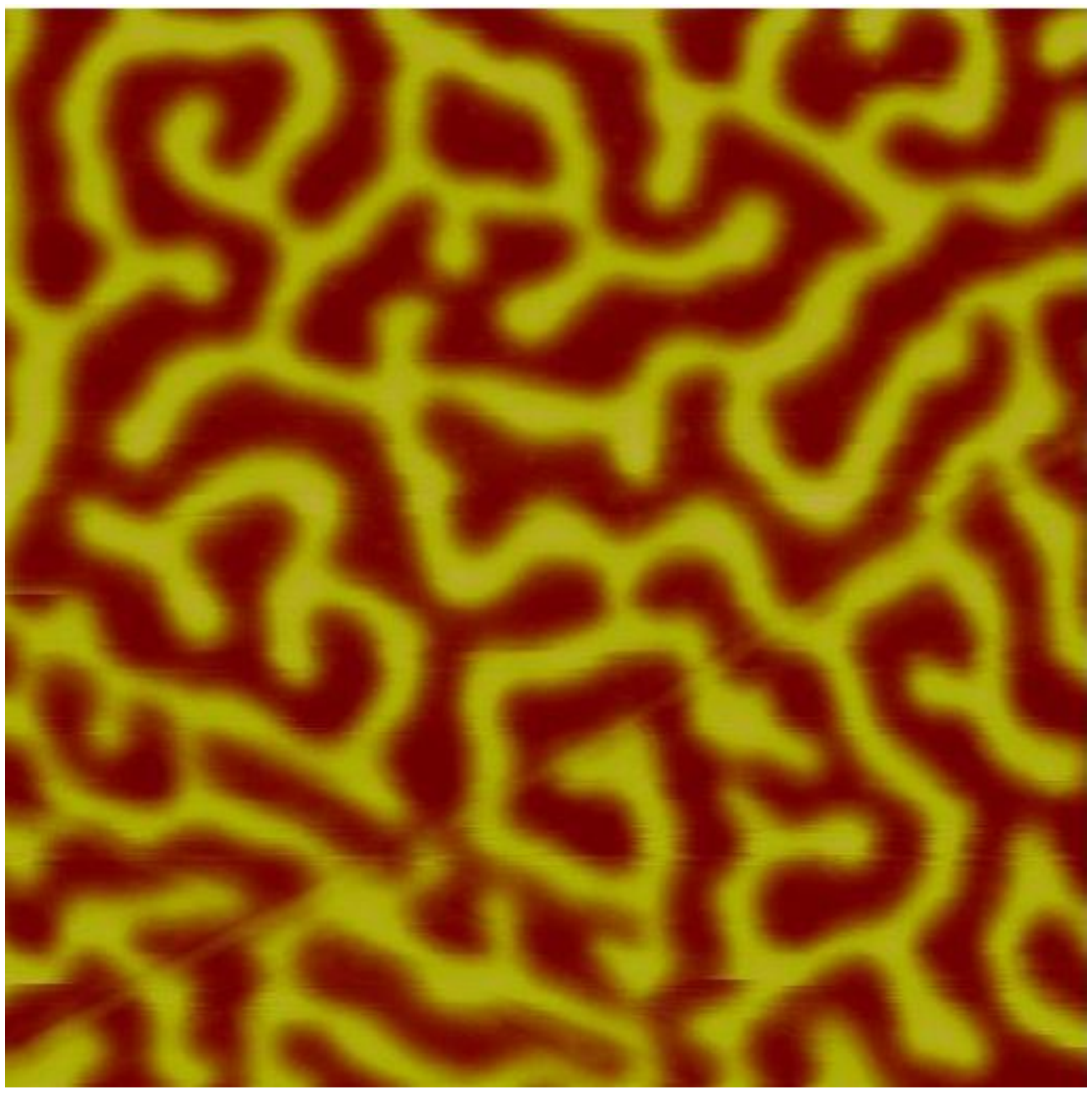

**Figure 5** (Color online) MFM image (5 μm × 5 μm) of sample B performed at $H = 0$ at room temperature after magnetizing it out of the plane of the sample.

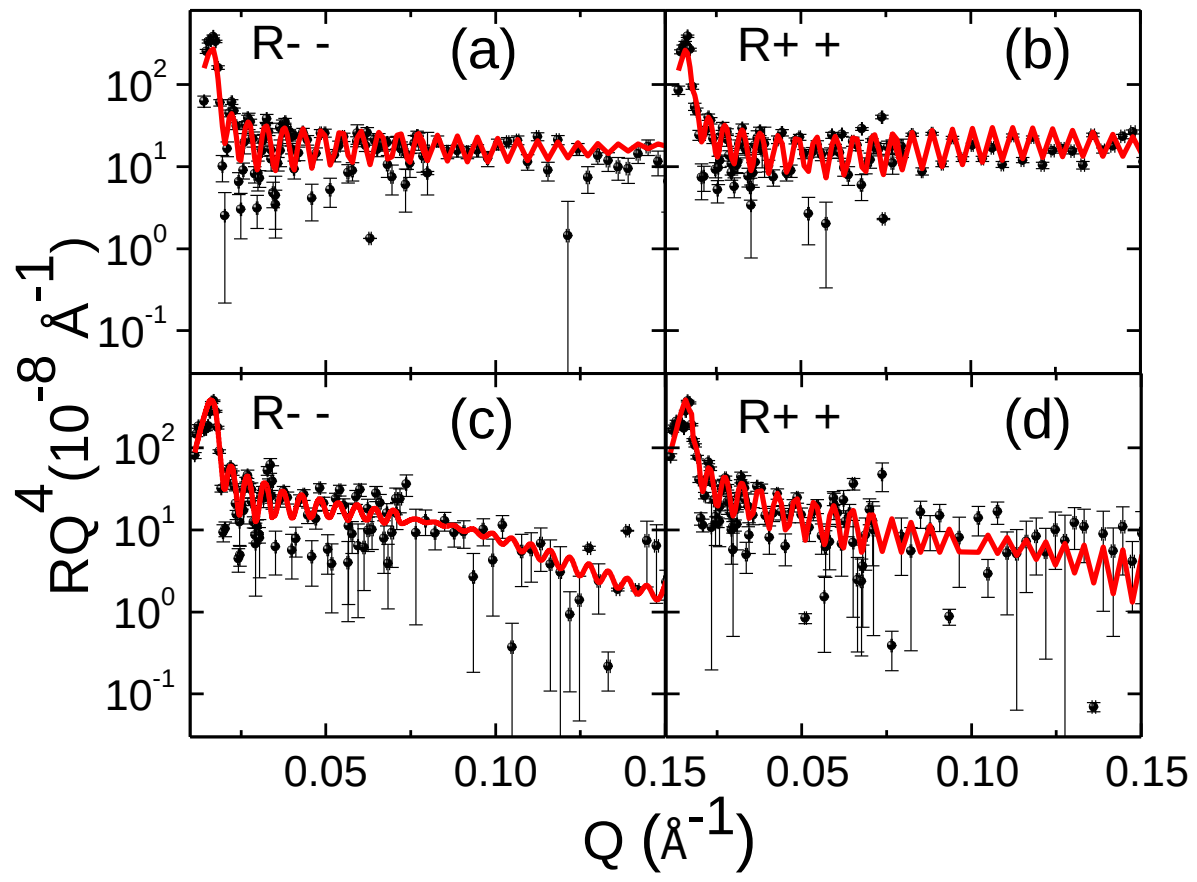

**Figure 6** (Color online) PNR using neutrons with (- -) and (+ +) incoming and outgoing polarization states with the sample in helium [(a) and (b)] and deuterium [(c) and (d)] in a 0.65 T magnetic field applied in the plane of sample A. Experimental data are black dots and the model fit is the red line.

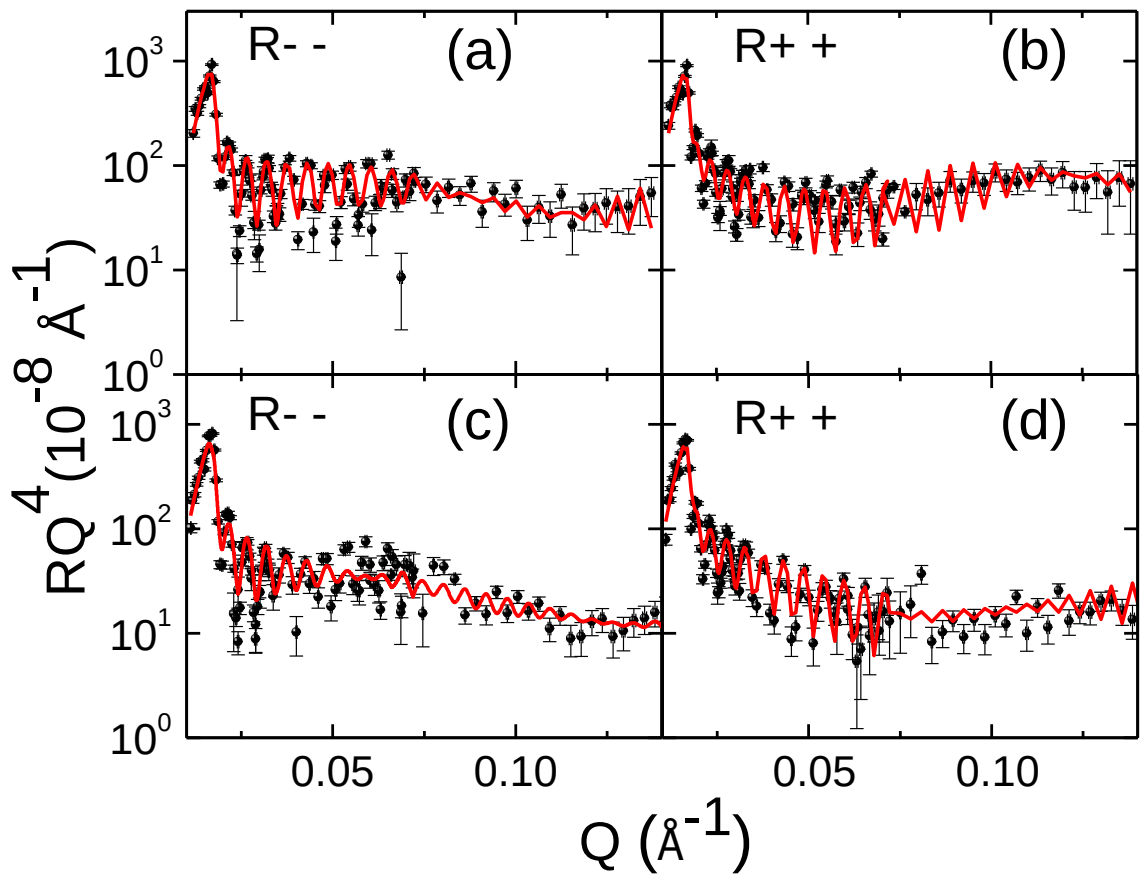

**Figure 7** (Color online) PNR using neutrons with (- -) and (+ +) incoming and outgoing polarization states with the sample in helium [(a) and (b)] and deuterium [(c) and (d)] in a 6 T magnetic field applied in the plane of sample A. Experimental data are black dots and the model fit is the red line.

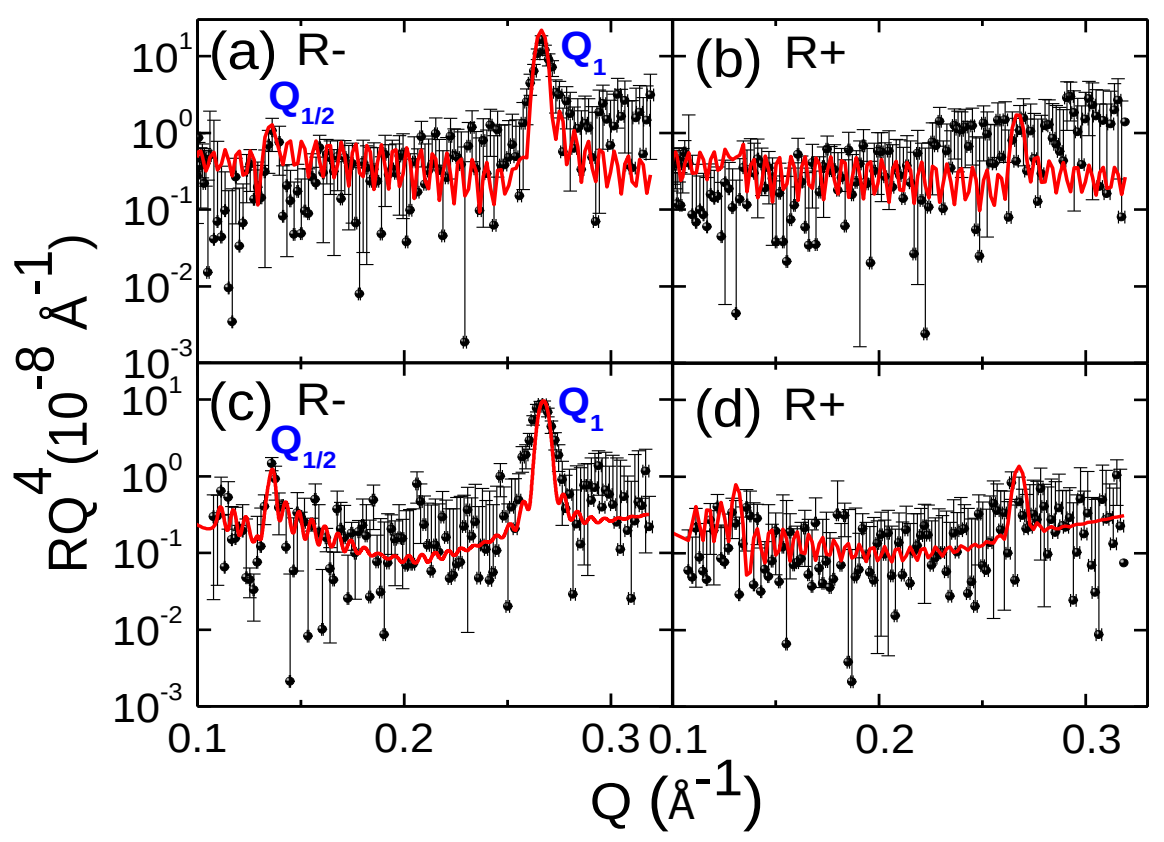

**Figure 8** (Color online) PNR using neutrons in the high-Q regime (Q>0.1 $Å^{-1}$) with (-) and (+) incoming polarization states with the sample in air [(a) and (b)] and deuterium [(c) and (d)] in a 0.65 T magnetic field applied in the plane of sample B. The positions of the single order ($Q_1$) and half-order magnetic peaks ($Q_{1/2}$) are indicated. Experimental data are black dots and model fit is red line.

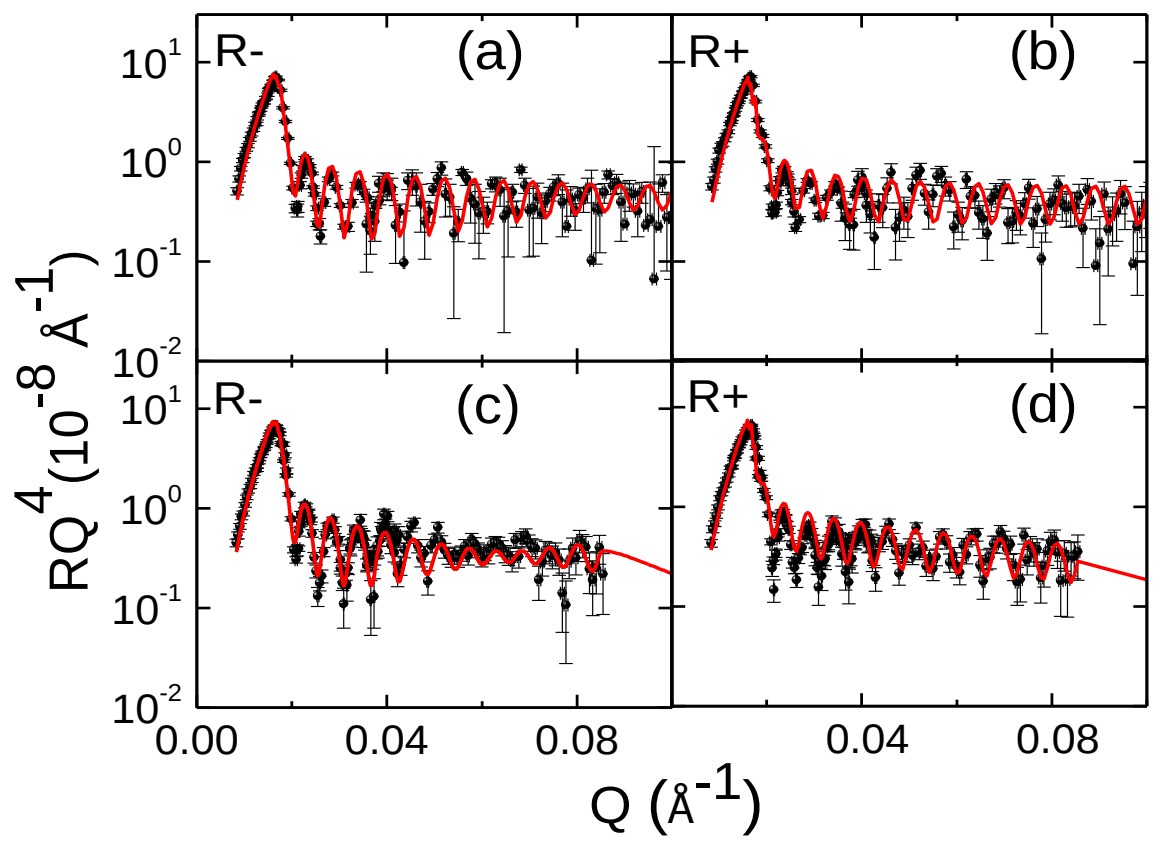

**Figure 9** (Color online) PNR using neutrons in the low Q regime (Q < 0.1 Å$^{-1}$) with (-) and (+) incoming polarization states with the sample in air [(a) and (b)] and deuterium [(c) and (d)] in a 0.65 T magnetic field applied in the plane of sample B in the low Q regime. Experimental data are black dots and model fit is red line.

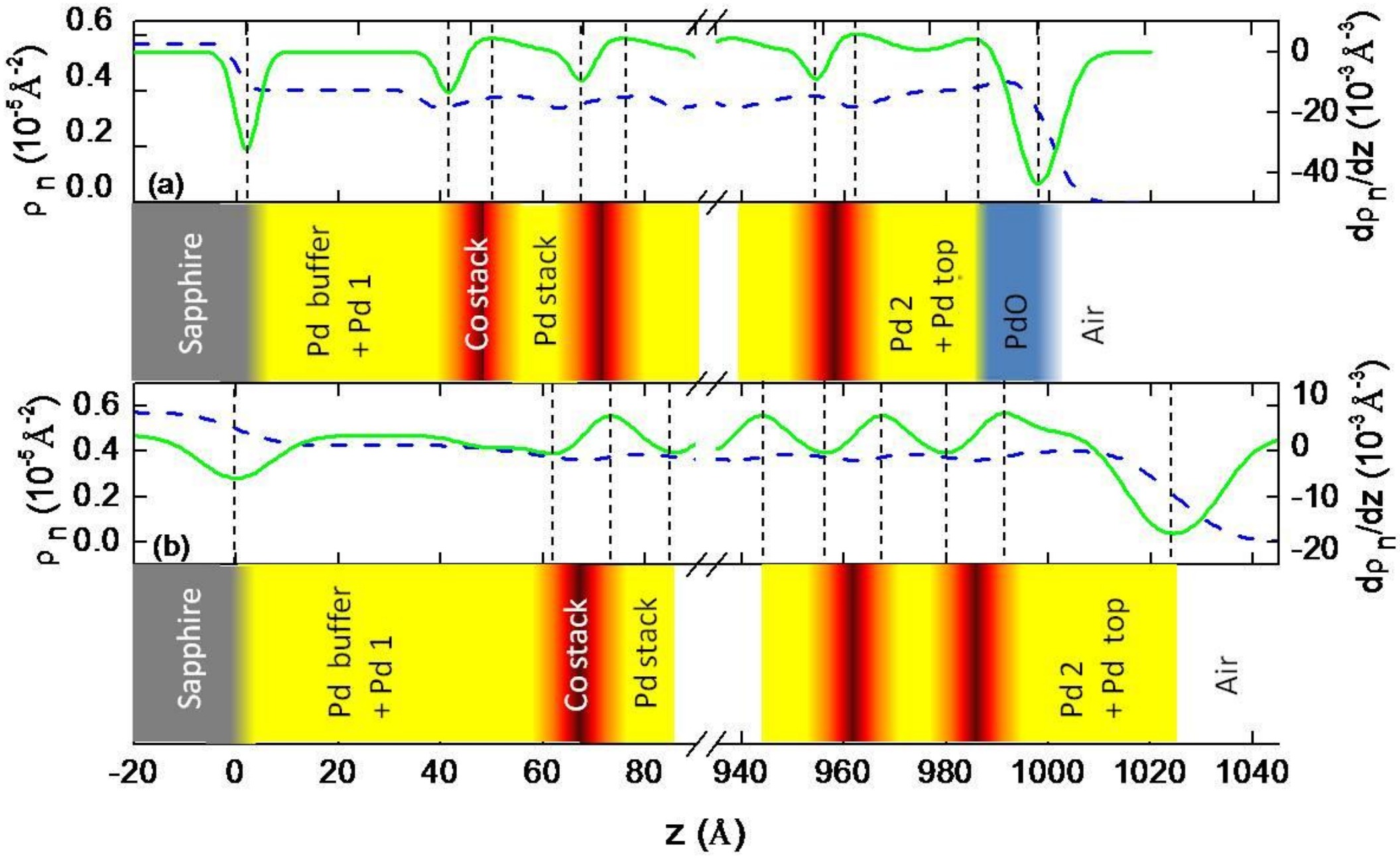

**Figure 10** (Color online) Nuclear SLD profiles (blue dashed curve) and its derivative (green solid curve for sample B (a) in air and (b) for the sample in deuterium. The vertical dotted lines indicate the positions of the interfaces. The corresponding sample profile is shown.

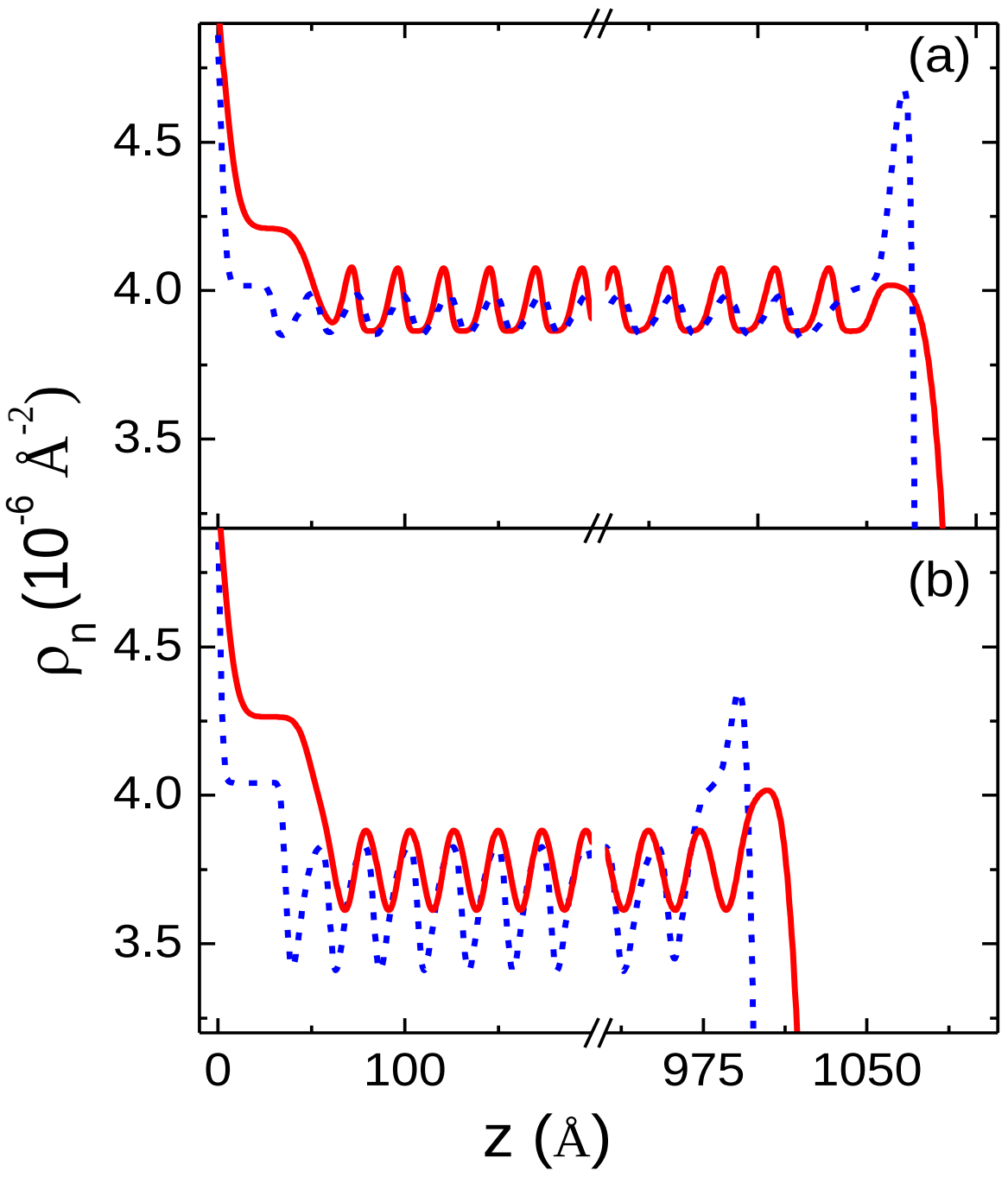

**Figure 11** (Color online) Nuclear SLD profile in air (helium) (blue dashed curve) and in deuterium (red solid curve) for (a) sample A and (b) sample B at 0.65 T.

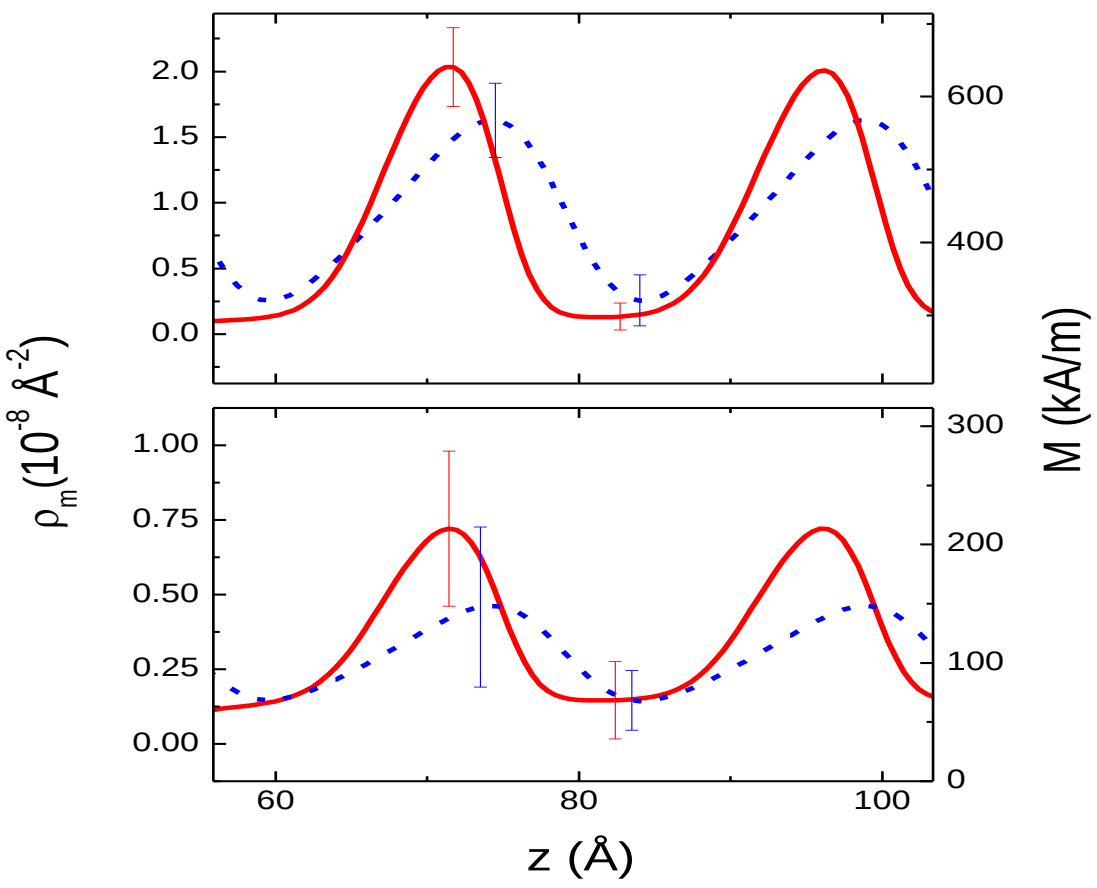

**Figure 12** (Color online) Magnetic SLD in air (blue dashed curve) and in $D_2$ (red solid curve) for sample A at (a) 6 T and (b) 0.65 T for two Co/Pd bilayer in the stack. The film-substrate interface is set at a thickness of zero. The equivalent magnetization, calculated by dividing $\rho_m$ by 2.853×10$^{-9}$ Å$^{-2}$/(10$^{-3}$ A/m).

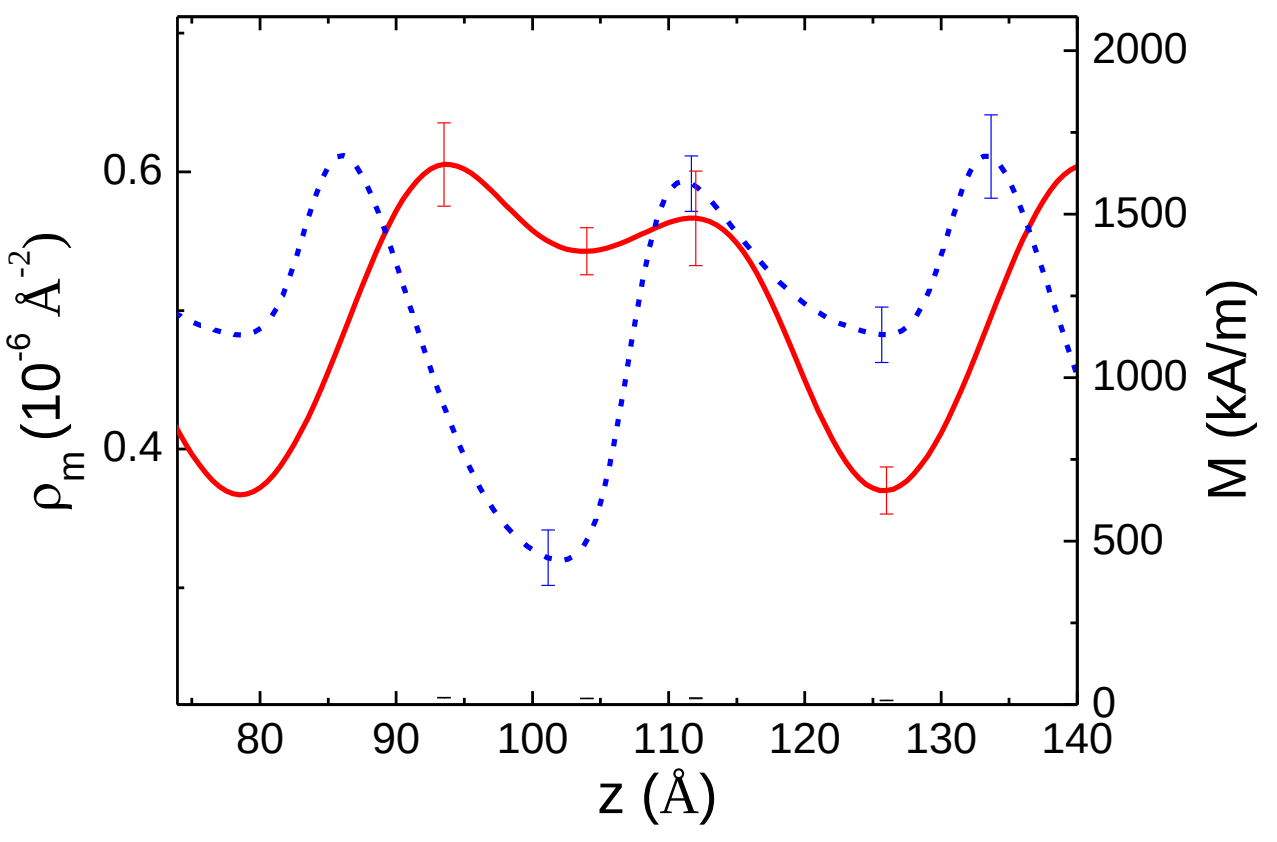

**Figure 13** (Color online) Magnetic SLD in air (blue dashed curve) and in $D_2$ (red solid curve) for sample B at 0.65 T for one Co/Pd bilayer in the stack. The film-substrate interface is set at a thickness of zero. The equivalent magnetization, calculated by dividing $\rho_m$ by 2.853×$10^{-9}$ $Å^{-2}$/($10^3$ A/m), is shown in the scale on the right.